\documentclass[sigconf]{acmart}
\usepackage{graphicx}
\usepackage{caption}
\usepackage{subcaption}
\usepackage{amsthm}
\usepackage{algorithm}
\usepackage{algorithmic}
\usepackage{graphicx}
\usepackage{subcaption}
\AtBeginDocument{%
  }

\setcopyright{acmlicensed}
\copyrightyear{2018}
\acmYear{2018}
\acmDOI{XXXXXXX.XXXXXXX}

\acmConference[Conference acronym 'XX]{Make sure to enter the correct
  conference title from your rights confirmation emai}{June 03--05,
  2018}{Woodstock, NY}
\acmISBN{978-1-4503-XXXX-X/18/06}

\begin{document}

\title{Efficient and Responsible Adaptation of Large Language Models for Robust Top-k Recommendations}

\author{Kirandeep Kaur}
\affiliation{%
  \institution{University of Washington}
  \city{Seattle}
  \state{WA}
  \country{USA}
}
\email{kaur13@cs.washington.edu}

\author{Chirag Shah}
\affiliation{%
  \institution{University of Washington}
  \city{Seattle}
  \state{WA}
  \country{USA}
}
\email{chirags@uw.edu}

\renewcommand{\shortauthors}{Trovato et al.}

\begin{abstract}
  Conventional recommendation systems (RSs) are typically optimized to enhance performance metrics uniformly across all training samples.
   This makes it hard for data-driven RSs to cater to a diverse set of users due to the varying properties of these users. The performance disparity among various populations can harm the model's robustness with respect to sub-populations.  While recent works have shown promising results in adapting large language models (LLMs) for recommendation to address hard samples, long user queries from millions of users can degrade the performance of LLMs and elevate costs, processing times and inference latency. This challenges the practical applicability of LLMs for recommendations. To address this, we propose a hybrid task allocation framework that utilizes the capabilities of both LLMs and traditional RSs. By adopting a two-phase approach to improve robustness to sub-populations, we promote a strategic assignment of tasks for efficient and responsible adaptation of LLMs. Our strategy works by first identifying the weak and inactive users that receive a suboptimal ranking performance by RSs. Next, we use an in-context learning approach for such users, wherein each user interaction history is contextualized as a distinct ranking task and given to an LLM. We test our hybrid framework by incorporating various recommendation algorithms -- collaborative filtering and learning-to-rank recommendation models -- and two LLMs -- both open and close-sourced. Our results on three real-world datasets show 
   improved robustness of RSs to sub-populations $(\approx12\%)$ and overall performance without disproportionately escalating costs. 
\end{abstract}  



\maketitle
\begin{figure}[t!]
\centering
\includegraphics[width=0.99\columnwidth]{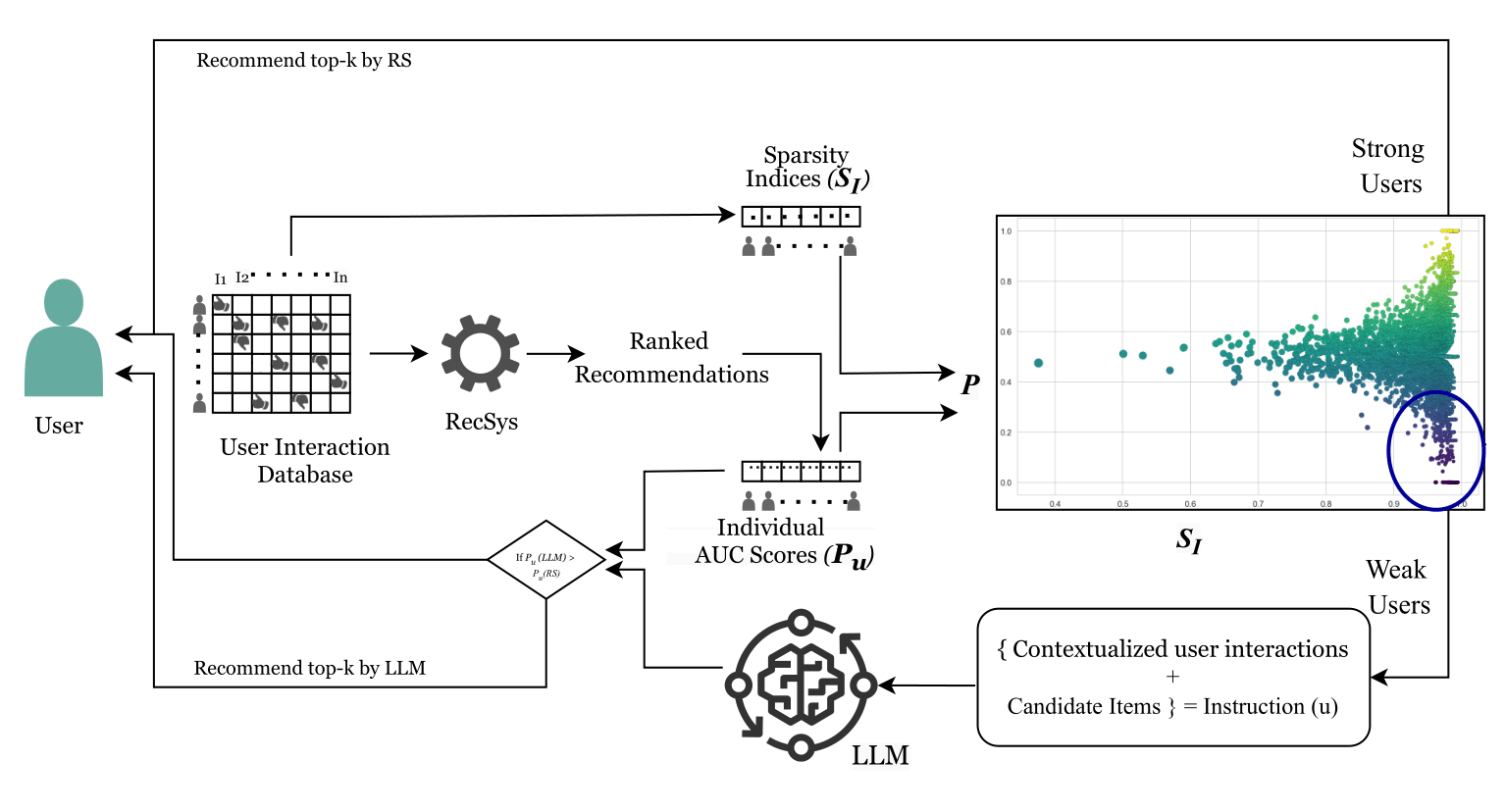}
    \caption{An overview of our framework that uses task allocation to adapt LLMs responsibly. We compute each user's sparsity index ($S_I$), evaluate recommendations retrieved from RS using performance metric ($P(u_m)$), and plot $P(u_m)$ against $S_I$. Interaction histories of highly sparse users with low $P(u_m)$ are contextualized and given to LLM for ranking. Strong users receive RS recommendations, while weak users get LLM recommendations if LLM outperforms RS.}
    \label{fig:mainlabel}
    \Description[Overall Framework]{Our framework employs a task allocation strategy for responsible adaptation of LLMs. Each user's sparsity index ($S_I$) is evaluated; traditional RS is trained on user interactions, generating recommendations with AUC scores ($P(u_m)$). Users' AUC scores are plotted against the sparsity index. Highly sparse users with low AUC scores receive contextualized interaction histories for LLM ranking; strong users receive RS recommendations, while weak users get LLM recommendations if they significantly outperform RS.}
\end{figure}

\section{Introduction}
\label{sec:introduction}
Recommendation systems (RSs) have become an integral part of numerous online platforms, assisting users in navigating vast amounts of content to relieve information overload~\cite{jacoby1984perspectives}. While Collaborative Filtering based RSs~\cite{papadakis2022collaborative} primarily rely on user-item interactions to predict users' preferences for certain candidate items, the utilization of language in recommendations has been prevalent for decades in hybrid and content-based recommenders, mainly through item descriptions and text-based reviews~\cite{lops2011content}. Furthermore, conversational recommenders~\cite{sun2018conversational} have highlighted language as a primary mechanism for allowing users to naturally and intuitively express their preferences~\cite{google_LLM+RS}. Deep recommendation models are trained under the Empirical Risk Minimization (ERM) framework that minimizes the loss function uniformly for all training samples. Such models, however, fail to cater to a diverse set of sub-populations, affecting robustness~\cite{geifman2017deep,wang2023exploring,chaney2018algorithmic,beutel2017beyond,yao2021measuring,zhang2021model,beutel2019fairness}. Empirical analysis conducted by~\citet{li2021user} shows that active users who have rated many items receive better recommendations on average than inactive users. This inadvertent disparity in recommendations requires careful scrutiny to ensure equitable recommendation experiences for all users~\cite {gunawardana2012evaluating}.

On the other hand, Large Language Models (LLMs) like GPT~\cite{achiam2023gpt}, LLaMA~\cite{touvron2023llama}, LaMDA~\cite{collins2021lamda}, Mixtral~\cite{jiang2024mixtral} can effectively analyze and interpret textual data, thus enabling a better understanding of user preferences. These foundation models demonstrate remarkable versatility, adeptly tackling various tasks across multiple domains~\cite{chen2023exploring,bansal2024llm,yang2023harnessing}. However, the field of recommendations is highly domain-specific and requires in-domain knowledge. Consequently, many researchers have sought to adapt LLMs for recommendation tasks~\cite{mohanty2023recommendation,huang2024foundation,fan2023recommender,lin2023can}. Authors in~\cite{lin2023can} outline four key stages in integrating LLMs into the recommendation pipeline: user interaction, feature encoding, feature engineering, and scoring/ranking. The purpose of using LLMs as a ranking function aligns closely with general-purpose recommendation models. The transition from traditional library-based book searches to evaluating various products, job applicants, opinions, and potential romantic partners signifies an important societal transformation, emphasizing the considerable responsibility incumbent upon ranking systems~\cite{singh2018}. Existing works that deploy LLMs for ranking~\cite{google_LLM+RS,hou2024large,xu2024prompting,wang2023multiple,ghosh2023jobrecogpt,zhang2023agentcf,tamber2023scaling,wang2023drdt,yue2023llamarec} have proven excellence of LLMs as zero-shot or few-shot re-rankers demonstrating their capabilities in re-ranking with frozen parameters. These works use traditional RSs as candidate item retrieval models to limit the candidate items that need to be ranked by LLM due to a limited context window. Furthermore,~\citet{xu2024prompting,hou2024large} interpret user interaction histories as prompts for LLMs and show that LLMs perform well only when the interaction length is up to a few items, demonstrating the ability of LLMs for (near) cold-start users. Since adapting LLMs can raise concerns around economic and efficiency factors, most of these works train RS on entire datasets but randomly sample interaction histories of some users to evaluate the performance of LLMs, questioning the generalizability of results for all users. This leads us to two important research questions. 

\begin{itemize}
    \item \textbf{RQ1}: Though LLMs have shown remarkable ranking performance even in zero-shot settings, how can we reduce the high costs associated with adapting LLMs to support practical applicability?
    \item \textbf{RQ2}: Conventional recommendation systems are cost-effective and can perform well on most users, as shown by previous works; how can we prevent performance degradation on sub-populations? 
\end{itemize}

To address these RQs, we propose a task allocation strategy that leverages LLM and RS's capabilities in a hybrid framework (Fig.~\ref{fig:mainlabel}). Our strategy operates in two phases based on the responsible and strategic selection of tasks for the cost-effective usage of LLMs. First, we identify the users with highly sparse interaction histories on whom the ranking performance of RS is below a certain threshold $t_p$. All such users are termed as weak users. In the second phase, interaction histories of weak users are contextualized using in-context learning to demonstrate user preferences as instruction inputs for LLM. While the strong users receive the final recommendations retrieved by RS, weak users receive the recommendations ranked by LLM if the quality of the ranked list is better than the RS. We test our framework based on collaborative filtering and learning-to-rank recommendation models and our results show the efficacy of our strategy, both with open-source as well as closed-source LLMs, in boosting the model robustness to sub-population and data sparsity and improving the quality of recommendations. For reproducibility and to support research community, our code is available on \hyperlink{https://anonymous.4open.science/r/resp-llmsRS/}{https://anonymous.4open.science/r/resp-llmsRS/}.
In short, the following are our contributions in this paper.

\begin{itemize}
    \item We introduce a {\bf novel hybrid task allocation strategy} that combines the strengths of LLMs and traditional RSs to improve robustness to subpopulations and data sparsity.
    \item Our {\bf unique method} for pinpointing weak users based upon two criteria (user activity and the received recommendation quality below a set threshold) facilitates interventions using LLMs for equitable recommendations.
    \item Our {\bf proposed framework improves the robustness of traditional recommendation models} by reducing weak user count, enhancing recommendation quality, and addressing high costs associated with adapting LLMs.
    \item Our {\bf experiments, both on closed-source and open-source LLMs}, show the efficacy of our framework in improving the model robustness to sub-populations by $(\approx 12\%)$ for varying levels of sparsity and reducing the count of weak users significantly.
\end{itemize}

\section{Related Work}
\label{sec:related_work}
Robustness in machine learning (ML) targets developing models capable of withstanding the challenges posed by imperfect data in diverse forms~\cite{zhang2019building}. Within the paradigm of recommendations, some existing works developed models resilient to shifts in popularity distribution~\cite{zhang2023invariant, zhang2024robust,10.1145/3583780.3615492}, distribution disparity in train and test datasets~\cite{yang2023generic,wang2024distributionally}, adversarial and data poisoning attacks~\cite{jia2023pore,burke2015robust,wu2023influence,tang2019adversarial,wu2021fight}. Our work aims to tackle the recommendation model's robustness to data sparsity~\cite{song2022data} and sub-populations~\cite{ovaisi2022rgrecsys}. 

In their research,~\citet{li2021user} illustrated that RSs excel in catering to active users but fall short in meeting the overall needs of inactive ones. To address this inequality, they proposed a re-ranking technique that reduced the disparity among active and inactive users. Their results depict that such post-processing techniques~\cite{DBLP:journals/corr/YangS16a,DBLP:journals/corr/ZehlikeB0HMB17,DBLP:journals/corr/CelisSV17} can either harm the average performance on advantaged users to reduce the disparity or reduce the overall utility of models. Though the in-processing techniques~\cite{pmlr-v81-kamishima18a, DBLP:journals/corr/abs-1805-08716,DBLP:journals/corr/YaoH17} for improving equitable recommendations across various sub-populations can tackle fairness-utility trade-offs, simply adding regularizer term results in sub-optimal performance~\cite{wang2023survey}. Most of these works have shown disparity and evaluated existing models by grouping users based on their activity, demographics, and preferences. Similarly,~\citet{wen2022distributionally} developed a Streaming-Distributionally Robust Optimization (S-DRO) framework to enhance performance across user subgroups, particularly by accommodating their preferences for popular items. Different from these, our work first builds upon the existing literature that elicits the issue of performance disparities among active and inactive users and then indicates that though inactive users receive lower-quality recommendations on average, this degradation only affects a subset of inactive users rather than all inactive users. Unlike these works, our framework identifies weak users— inactive individuals whose preferences traditional recommendation systems struggle to capture effectively.

Many researchers have turned to LLMs to address some of these problems because, in recent years, LLMs have proven to be excellent re-rankers and have often outperformed existing SOTA recommendation models in zero-shot and few-shot settings without requiring fine-tuning. For example,~\citet{Gao2023ChatRECTI} proposed an enhanced recommender system that integrates ChatGPT with traditional RS by synthesizing user-item history, profiles, queries, and dialogue to provide personalized explanations to the recommendations through iterative refinement based on user feedback. AgentCF~\cite{zhang2023agentcf}, designed to rank items for users, involves treating users and items as agents and optimizing their interactions collaboratively. While user agents capture user preferences, item agents reflect item characteristics and potential adopters’ preferences. They used collaborative memory-based optimization to ensure agents align better with real-world behaviours. While the retrieval-ranker framework in~\cite{wang2023multiple} remains similar to previous works, authors generate instructions with key values obtained from both users (e.g., gender, age, occupation) and items (e.g., title, rating, category). 

Despite the excellence of LLMs as ranking agents, adapting LLMs can involve processing lengthy queries containing numerous interactions from millions of users. Furthermore, each query can raise various economic and latency concerns. Thus, all these works randomly select a few users from the original datasets to evaluate the performance of LLMs. In practice, this user base can involve many more users, which questions the practical applicability of large models for recommendations. However, some recent studies have shown the efficacy of large language models (LLMs) as re-ranking agents to cater to queries with shorter interaction histories compared to lengthy instructions that constitute hundreds of interactions. 

For example,~\citet{hou2024large} trained recommendation systems to generate candidate item sets and then used user-item interactions to develop instructions. The authors sorted users' rating histories based on timestamps and used in-context learning to design recency-focused prompts. They prompted LLMs to re-rank the candidate items retrieved by the recommendation systems. Their analysis showed decreased performance of LLMs if the candidate item set had more than 20 items. ProLLM4Rec \cite{xu2024prompting} adopted a unified framework for prompting LLMs for recommendation. The authors integrated existing recommendation systems and works that use LLMs for recommendations within a single framework. They provided a detailed comparison of the capabilities of LLMs and recommendation systems. Their empirical analysis showed that while state-of-the-art sequential recommendation models like SASRec \cite{kang2018self} improve with a growing number of interactions, LLMs start to perform worse when the number of interactions grows. Furthermore, both of these works sampled some users to evaluate the performance of LLMs due to the high adaptation costs. To investigate the effectiveness of various prompting strategies \citet{google_LLM+RS} focused on a (near) cold-start scenario where minimal interaction data is available. They used various prompting techniques to provide a natural language summary of preferences to enhance user satisfaction by offering a personalized experience. By exploiting rich positive and negative descriptive content and item preferences within a unified framework, they compared the efficacy of prompting paradigms with large language models against collaborative filtering baselines that rely solely on item ratings.

In summary, past works suggest that despite the high costs associated with adapting LLMs for recommendations, these models can outperform existing recommendation models significantly. Moreover, we acknowledge that the literature shows the contrasting capabilities of both RSs and LLMs -- RSs fail to perform well on inactive users due to sparse interaction vectors, and in contrast, LLMs can be prompted to cater to inactive users in near cold-start settings without requiring any fine-tuning.

Building upon these crucial insights, our framework first aims to identify the weak users for whom RS finds it hard to capture their preferences accurately. We then use in-context learning to prompt LLMs to generate recommendations for such users. While past works like ProLLM4Rec by~\cite{xu2024prompting}, dynamic reflection with
divergent thinking within a retriever-reranked by~\cite{wang2023drdt}, recency-focused prompting by~\cite{hou2024large} and aligning ChatGPT with conventional ranking techniques such as point-wise, pair-wise, and list-wise ranking by~\cite{dai2023uncovering} are all different techniques to design prompts with different variations, our main contribution lies in the responsible task allocation within recommendation systems and all such techniques can be used within our framework for designing prompts. In the next section, we discuss our methodology in detail.

\section{Methodology}
\label{sec:methodology}
We begin here by providing a formal definition of the existing problem. We then discuss our framework, which adopts a hybrid structure by leveraging the capabilities of both traditional RSs and LLMs. For this, we first identify users for whom RSs do not perform well and then leverage LLMs for these users to demonstrate user preferences using in-context learning.

\subsection{Problem Formulation}
Consider a recommendation dataset $\mathcal{D}$ with $k$ data points. Let $U = \{u_1,u_2,\dots,u_M\}$ be the set of users and $|U| = M$ represents the number of users in $\mathcal{D}$. Let $I = \{i_1,i_2,\dots,i_N\}$ be the set of all the items and $|I| = N$ represents the number of items in $\mathcal{D}$.

\begin{align}
\mathcal{D} = \{(u_m,i_n,r_{mn}): m=1,2,\dots,M; n = 1,2,\dots,N\}
\end{align}

Here, the triplet $d_{mn} = (u_m,i_n,r_{mn})$  represents one data point where a user $u_m$ provided a rating of $r_{mn}$ to an item $i_n$. Now, if a user $u_m$ has rated a set of items, then let $[r_{mn}]_{n=1}^{N}$ denote the rating vector consisting of explicit rating values ranging from $1$ to $5$ if a user provided a rating and $0$ otherwise. Additionally, $\theta^r$ represents the conventional recommendation model. The first step to solving the problem includes determining different criteria to categorize a user as weak. This includes ranking users based on the RS performance on each one of them. 
Then, the goal is to understand user characteristics to categorize extremely weak users. 
For each weak user, we contextualize interaction history as a distinct recommendation task and finally allocate these tasks to LLM. 

\subsection{Identifying Weak Users}
We consider two criteria for identifying weak users for recommendation model $\theta^r$. First, given $K$ users and their associated rating vectors, we evaluate how well the model could rank the relevant user items, often termed as \emph{positive items} above the irrelevant or \emph{negative items}. Let $r$ denote the rank of the relevant item, and $r'$ be the rank of irrelevant items. Then,

\begin{align}
    \delta(r < r')
\end{align}

\noindent
denotes an indicator function that outputs one if the rank of the relevant item $r$ is higher than that of the irrelevant item $r'$. Let $N$ denote the total number of items and $|R|$ be the set of all relevant items.  Then, similar to~\citet{rendle2012bpr}, we use AUC measure to evaluate how hard it was for $\theta^r$ to rank items preferred by a certain user, given by

\begin{align}
    \mathrm{\mathcal{P}(u)} &= \frac{1}{|R|(N-|R|)} \sum_{r \in R} \sum_{r' \in \{1, \dots, N\} \setminus R} \delta(r < r')
    \label{eq:p(u)}
\end{align}

Here, $|R|(n-|R|)$ denotes all possible pairs of relevant and irrelevant items.

We acknowledge that various metrics like NDCG, F1, precision and recall have been used to measure the quality of ranking ability of recommendation models. However, these metrics place significant importance on the outcomes of the top-k items in the list and completely ignore the tail. For identifying weak users, we require a metric consistent under sampling i.e. if a recommendation model tends to give better recommendations than another on average across all the data, it should still tend to do so even if we only look at a smaller part of the data. The aim of our framework is a clear task distribution. The performance of top-k metrics varies with $k$, and this might raise uncertainty as $k$ varies with varying users and platforms. Nevertheless, AUC is the only metric which remains consistent under-sampling, and as $k$ reduces, all top-k metrics collapse to AUC. For more details, we refer the readers to~\cite{krichene2020sampled}. 


Past works~\cite{li2021user} have shown that \emph{active} users that provide more ratings receive better recommendations than the \emph{inactive} users on average. However, only a few inactive users might receive irrelevant recommendations individually (Fig.~\ref{fig:scatterplot}). Thus, we evaluate each user's activity. Let a user $u$ rated $|R|$ items out of a total of $N$ items. Then sparsity index $\mathcal{S}_I$ associated with a given user $u$ can be calculated as:

\begin{align}
    \mathcal{S}_I(u) = \frac{|R|}{N} 
    \label{eq:s(u)}
\end{align}

If this value falls above a certain threshold $t_s$, the user is considered as inactive. Combining with the weak user identification, we obtain,

\begin{definition}
    Given dataset $\mathcal{D}$ and a recommendation model $\theta^r$, we say that a user $u_m$ is extremely weak if the likelihood of $\theta^r$ being able to rank the relevant items above the irrelevant items is below $t_p$ and the rating vector $[r_{mn}]$ has extremely high sparsity. i.e., above $t_s$
\end{definition}
\begin{align}
    \mathcal{P}(u_m) \leq t_p\hspace{1em} \&\&\hspace{1em}  \mathcal{S}_I(u) > t_s
\label{def1}
\end{align}

It is important to note that a higher AUC value implies better performance, and the value always lies between 0 to 1.  Further, we use $t_s = avg(\mathcal{S}_I(D))$, the average sparsity of all users in $\mathcal{D}$ i.e.,\\ $avg(\mathcal{S}_I(D)) = 1/m * \sum_{j=1}^m \mathcal{S}_I(u_j)$ for determining this threshold.

\subsection{Designing Natural Language Instructions for Ranking} 
Closest to our work,~\citet{hou2024large} formalized the recommendation problem as a \emph{conditional} ranking task considering sequential interaction histories as conditions and uses the items retrieved by traditional RS as $candidate$ items. While we aim to design the conditional ranking tasks, our approach differs significantly from theirs as instead of using LLMs as a re-ranking agent for all users; we instruct LLM with the preferences of weak users preferences (\emph{sorted in descending order of decreased preference}). This technique is detailed below. 

For each user, we use in-context learning to instruct LLM about user preferences \emph{conditions} and assign the task of ranking the \emph{candidate} items. For a user $u$, let $\mathcal{H}_{u} = \{i_1,i_2,\dots,i_n\}$ depict the user interaction histories sorted in decreasing order of preference and $\mathcal{C}_{u} = \{i_1,i_2,\dots,i_j\}$ be the candidate items to be ranked. Then, each instruction can be generated as a sum of conditions and candidate items, i.e.,

\begin{align}
    \mathcal{I}_{u} = \mathcal{H}_{u} + \mathcal{C}_{u} 
    \label{eq:I(u)}
\end{align}

\textbf{In-context learning:} We use in-context learning to provide a demonstration of the user preferences to LLM using certain examples. As suggested by~\citet{hou2024large}, providing examples of other users may introduce extra noise if a user has different preferences. Therefore, we sort every weak user's preferences based on explicit user ratings. For example, \emph{"User $\{user\_id\}$ liked the following movies in decreasing order of preference where the topmost item is the most preferred one: 1. Harry Potter, 2. Jurassic Park \dots"}. This forms the condition part of the instruction. We then select items which served as test items for recommendation models as candidate items and instruct LLM to rank them in decreasing order of preference as \emph{"Now, rank the following items in decreasing order of preference such that the top most movie should be the most preferred one: Multiplicity, Dune \dots  "}.

\begin{figure}
    \centering
    \includegraphics[width=0.45\textwidth]{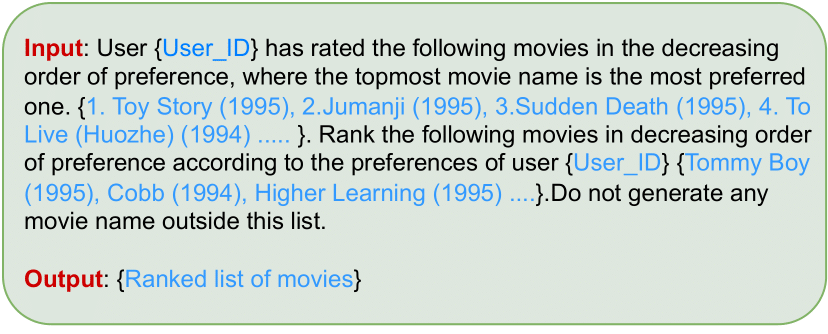}
    \caption{Instruction template for contextualizing interaction histories of weak users.}
    \label{fig:template1}
    \Description[Instruction template]{Instruction template for contextualizing interaction histories of weak users.}
\end{figure}

It is important to note that while the presentation order in conditions plays a significant role in demonstrating user preferences to LLM, we deliberately shuffle the candidate items to test the ability of LLM to rank correctly. Since LLMs can generate items out of the set, we specially instruct to restrict recommendations to the candidate set. Fig.~\ref{fig:template1} shows the final template of the instruction given to LLM for a particular user. We use the same template for all identified weak users to contextualize their past interactions into a ranking task.

\subsection{Our Framework}
This section discusses the workflow adopted by our framework as depicted in Fig.~\ref{fig:mainlabel} and corresponding algorithm~\ref{alg:framework}. Initially, the model takes input as the training $\mathcal{D}_{train}$ and test dataset $\mathcal{D}_{test}$, a set of users $\mathcal{U}$,  a recommendation model $\theta^r$, large language model $\theta^l$ and two thresholds: sparsity threshold $t_s$ and performance threshold $t_p$ which depict the minimum sparsity and performance values for user to be classified as strong user. It is important to note that splitting data will not yield a mutually exclusive set of users in both sets, but item ratings for each user in $\mathcal{D}_{train}$ will differ from those in $\mathcal{D}_{test}$.

The algorithm begins by training the recommendation model $\theta^{r}$ on the training set $\mathcal{D}_{train}$ and provides ranked items for all users. Using $\mathcal{D}_{test}$, we test the ranking ability of the model for each user by evaluating $\mathcal{P}(u_m)$ using Eq.~\ref{eq:p(u)}. Further, each user is also assigned a sparsity score $\mathcal{S}_I(u)$ evaluated using Eq.~\ref{eq:s(u)}. If $\mathcal{P}(u)$ has a value less than $t_p$ and the sparsity index $\mathcal{S}_I(u)$ for a particular user falls below $t_s$, the user is termed as a weak user. While previous works have shown that, on average, inactive users receive poor performance, we pinpoint weak users by evaluating both the sparsity and performance.

For all such weak users, we convert rating histories from $\mathcal{D}_{train}$ as conditions $\mathcal{H}_{u}$ using in-context learning and use test items as candidate items $\mathcal{C}_u$ for testing purposes. However, in practice, these candidate items can be replaced by unobserved items. The final instructions are generated by combining conditions and candidate items as depicted by (Eq.\ref{eq:I(u)}). These instructions are given to the LLM, which provides a ranked list of items for each user. For all the strong users, the recommendations presented are the ones ranked by the conventional recommendation model. However, the weak users receive final ranked lists generated by the LLM.

\begin{algorithm}[t!]
\caption{Hybrid LLM-RecSys Algorithm for Ranking}
\label{alg:framework}
\begin{algorithmic}[1]
    \STATE \textbf{Input}: $\mathcal{D}_{train}$: training dataset; $\mathcal{D}_{test}$: test dataset; $\mathcal{U}$: set of users; $\mathcal{S}_I$: Sparsity index for all users; $\theta^r$: recommendation algorithm; $\theta^l$: large language model, $t_{s}$: sparsity threshold, $t_{p}$: performance threshold.
    \STATE \textbf{Output}: $ranked\_pred_{strong}$: ranked lists of items for strong users,  $ranked\_pred_{weak}$: ranked lists of items for weak users.
    \STATE $ranked\_pred \leftarrow \theta^r(\mathcal{D}_{train}$)
    \FOR{ each user $u_{m} \in \mathcal{U}$}
        \STATE Calculate $\mathcal{P}(u_m)$ using Eq.~\ref{eq:p(u)}
        \STATE Calculate $\mathcal{S}(u_m)$ using Eq.~\ref{eq:s(u)}
        \IF{$\mathcal{P}(u_m) < t_{p}$ $\&\&$ $\mathcal{S}_I(u_m)$ }
            \STATE $\mathcal{U}_{weak} \leftarrow u_m$
        \ELSE
            \STATE $\mathcal{U}_{strong} \leftarrow u_m$
            \STATE $ranked\_pred_{strong} \leftarrow ranked\_pred[u_m]$
        \ENDIF
        
    \ENDFOR
    \FOR{each $u_{i} \in \mathcal{U}_{weak} $}
        \STATE Generate  instruction $\mathcal{I}_{u_i}$ using Eq. \ref{eq:I(u)}
        \STATE $ranked\_list_{u_{i}}$ = $\theta^{L}(\mathcal{I}_u)$
        \STATE $ranked\_list_{weak} \leftarrow ranked\_list_{u_{i}}$ 
    \ENDFOR
\end{algorithmic}
\end{algorithm}

\section{Experiments}
This section discusses our experimental setup with details of the datasets and models used, followed by the implementation details of all these models and various metrics used. We finally present empirical results and a comparative analysis of various recommendation models and LLMs.

\begin{table}[t!]
\centering
\caption{Datasets statistics}
\label{tab:dataset_statistics}
\resizebox{0.75\columnwidth}{!}{%
\begin{tabular}{c|c|c|c}
\hline\hline
                & \textbf{ML-1M}               & \textbf{ML-100k}      & \textbf{Book-Crossing}\\ \hline
\# Users        & \multicolumn{1}{c|}{6,041}   & 943    & 6,810                  \\
\# Items        & \multicolumn{1}{c|}{3,952}   & 1,682        & 9,135           \\
\# Interactions & \multicolumn{1}{c|}{1,000,209} & 100,000     &  114,426               \\
Sparsity        & \multicolumn{1}{c|}{95.81\%}  &  93.7\%       & 99.82\%            \\
Domain          & \multicolumn{1}{c|}{Movies}  & Movies     & Books              \\ \hline\hline
\end{tabular}%
}
\label{tab:my-table}
\end{table}

\subsection{Experimental Setup}

\subsubsection{Datasets.} To test the effectiveness of our framework, we conducted experiments on three real-world datasets: \textbf{ML-1M}\footnote{https://grouplens.org/datasets/movielens/1m/}, \textbf{ML100k}\footnote{https://grouplens.org/datasets/movielens/100k/}, and \textbf{Book-Crossing (B-C)}\footnote{http://www2.informatik.uni-freiburg.de/~cziegler/BX/}. Both ML100k and ML1M are movie-rating datasets, and book-crossing is a book-rating dataset. We select three datasets with varying levels of sparsity for evaluating robustness to data sparsity- ML100k has the least sparsity, and Book-Crossing has the highest sparsity (for exact values, refer to table~\ref{tab:dataset_statistics}). All these datasets have explicit user preference in the form of ratings ranging from $0-5$ for movie ratings and $0-10$ for book ratings dataset. We do not filter out users from ML1M and ML100k as each user has rated at least $20$ movies in both these datasets. For consistency, we filter out users with less than $20$ ratings from the Book-Crossing dataset. While both movie-ratings datasets have clustering based on sensitive attributes like age and gender, this paper aims to boost performance on all weak users irrespective of the sensitive features. Thus, following the protocol adopted by~\cite{li2021user}, we divided users based on their activity or the number of items rated. Any user who has rated items below a certain threshold $t_s$ is termed an inactive user, and all those above this threshold are active users. We calculated the number of items rated on an average by all the users and used this average value as a threshold; this threshold can always vary and be set to different values per application. Table~\ref{tab:dataset_statistics} presents the statistics of all three datasets.

\begin{figure*}[th!]
    \begin{minipage}[t]{1\textwidth}
        \centering
        \begin{subfigure}[t]{0.24\textwidth}
            \includegraphics[width=\textwidth]{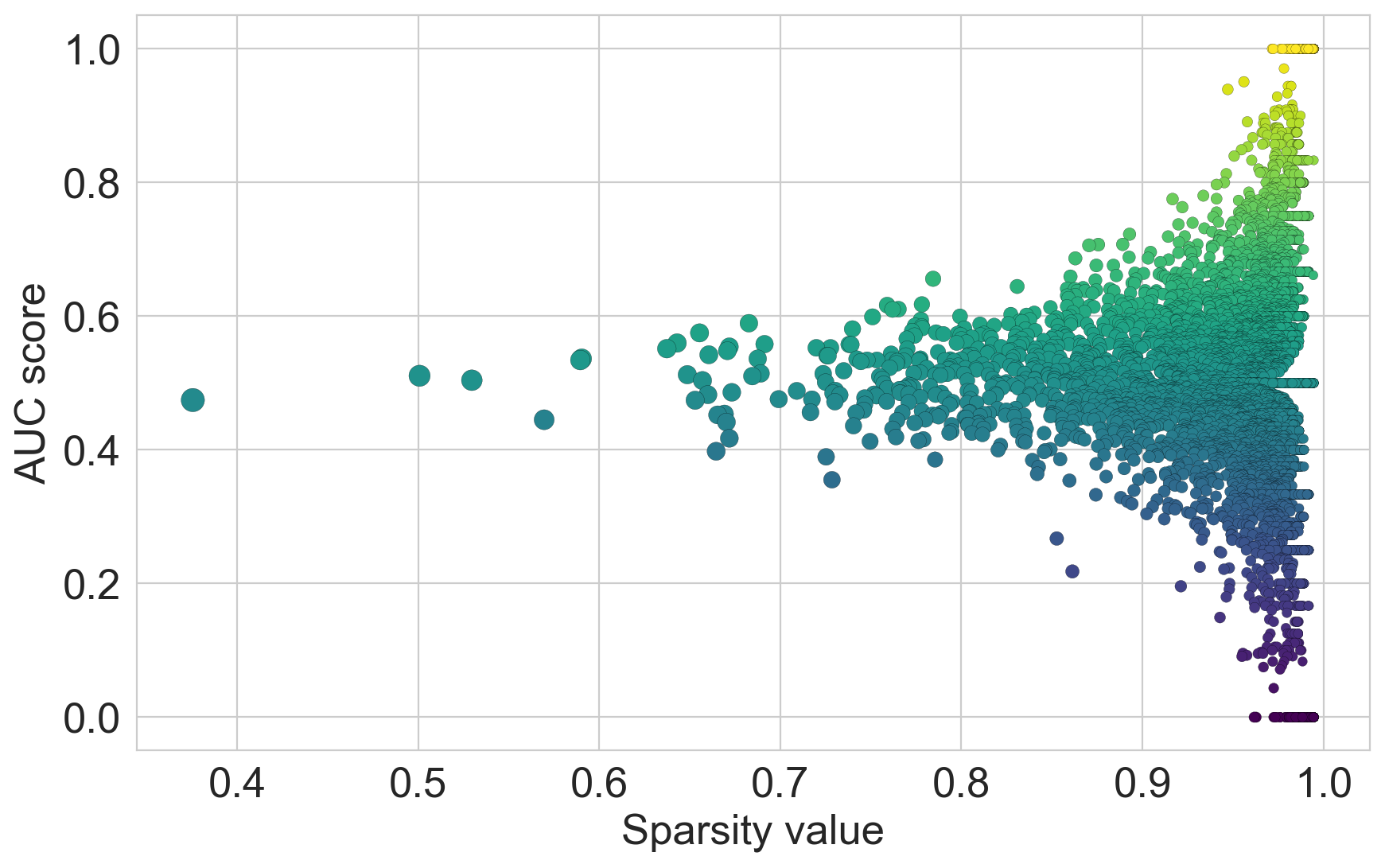}
            \caption{ItemKNN (ML1M)}
            \label{sfig:figure1}
        \end{subfigure}
        \begin{subfigure}[t]{0.24\textwidth}
            \includegraphics[width=\textwidth]{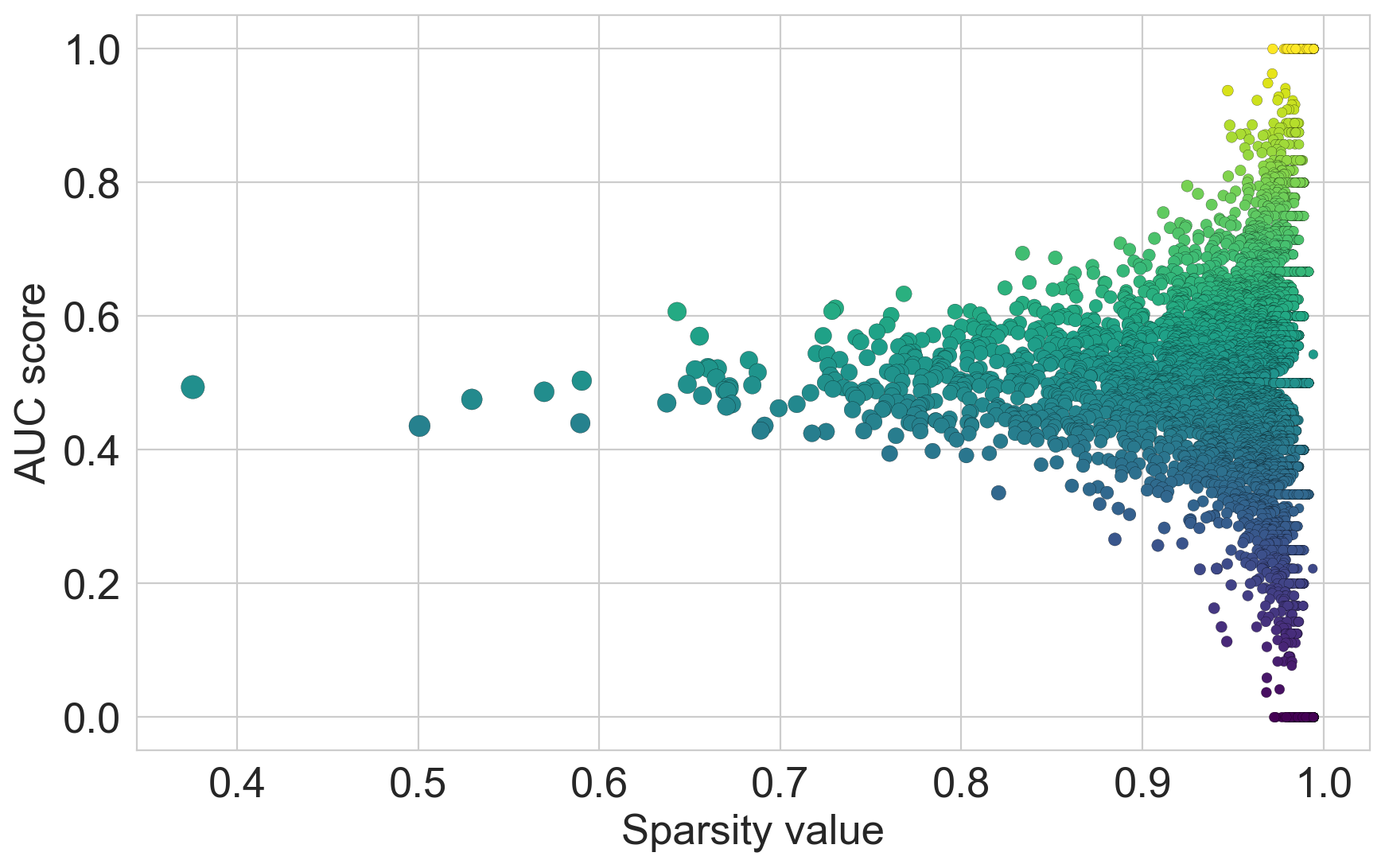}
            \caption{NCF (ML1M)}
            \label{ssfig:figure2}
        \end{subfigure}
        \begin{subfigure}[t]{0.24\textwidth}
            \includegraphics[width=\textwidth]{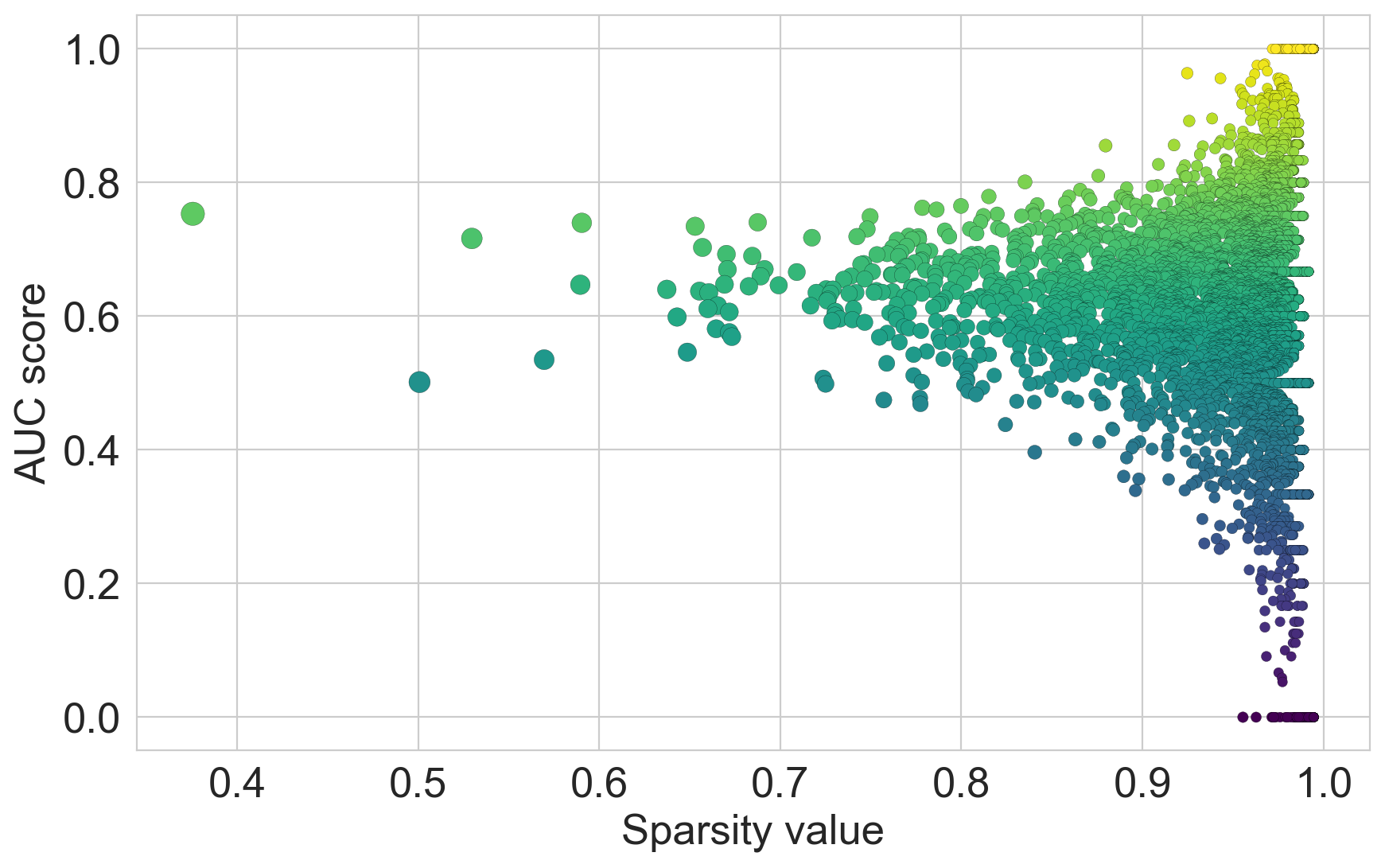}
            \caption{BPR (ML1M)}
            \label{sfig:figure3}
        \end{subfigure}
    \end{minipage}
    \begin{minipage}[t]{1\textwidth}
        \centering
        \begin{subfigure}[t]{0.24\textwidth}
            \includegraphics[width=\textwidth]{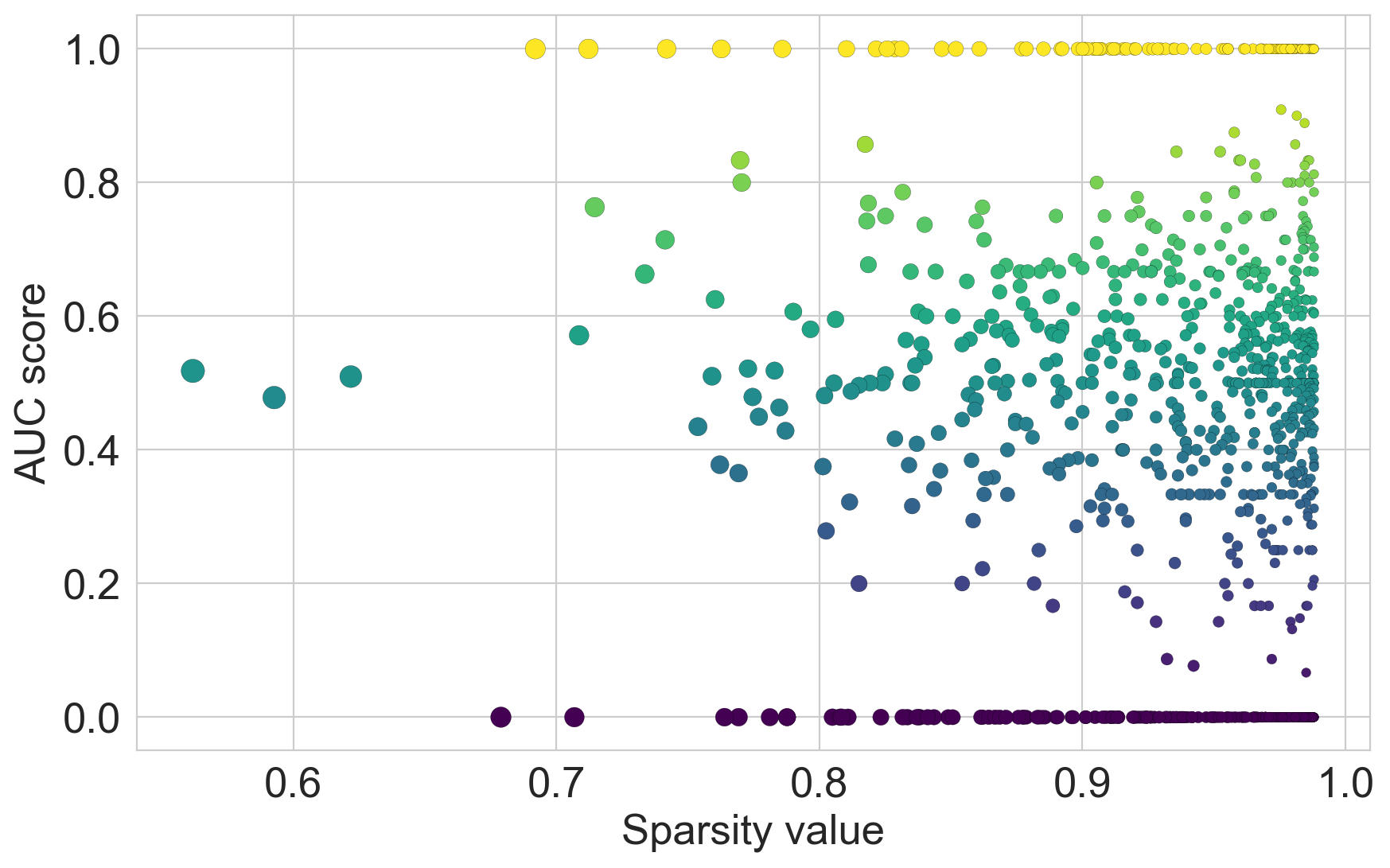}
            \caption{ItemKNN (ML100k)}
            \label{sfigg:figure4}
        \end{subfigure}
        \begin{subfigure}[t]{0.24\textwidth}
            \includegraphics[width=\textwidth]{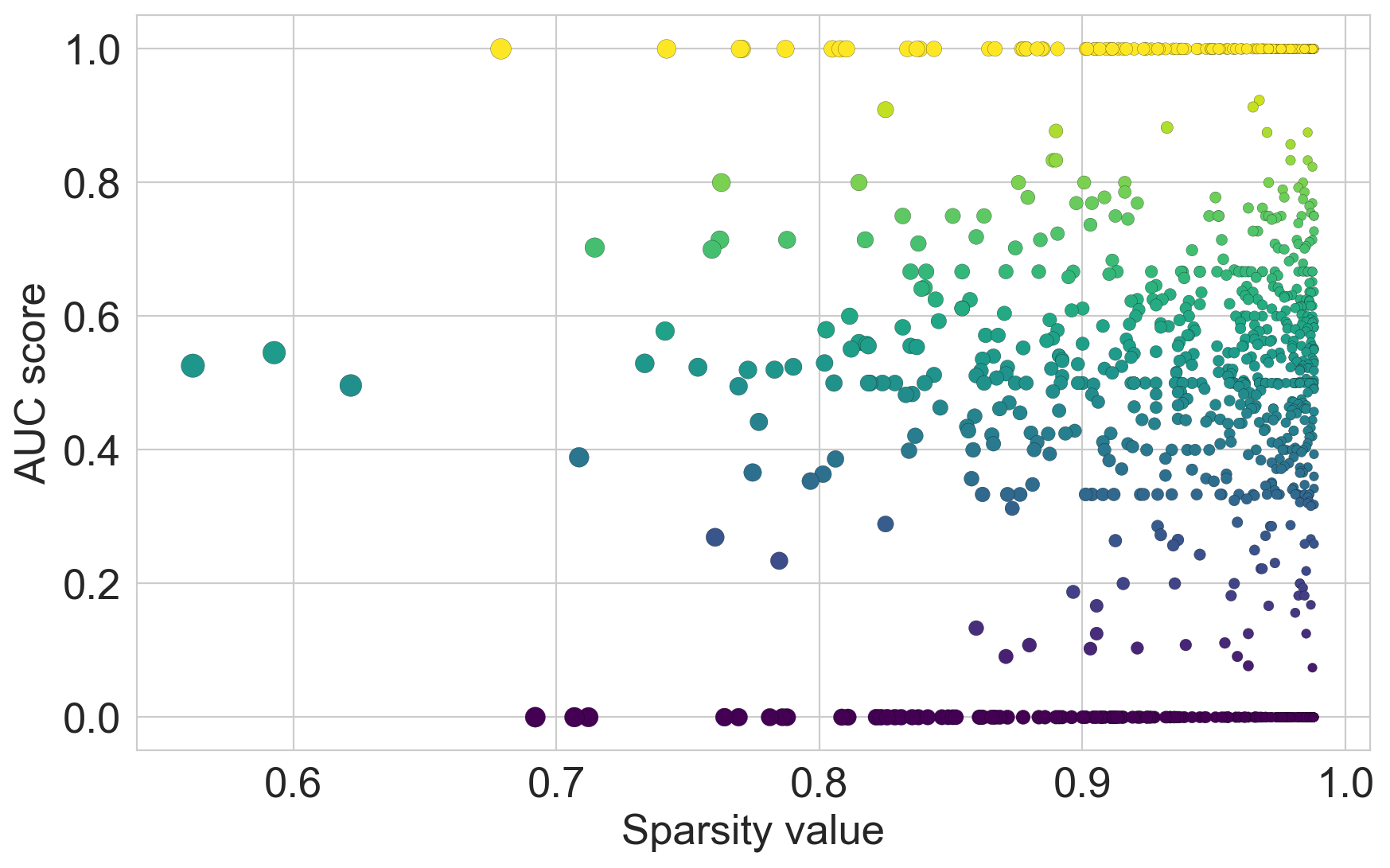}
            \caption{NCF (ML100k)}
            \label{sfig:figure2}
        \end{subfigure}
        \begin{subfigure}[t]{0.24\textwidth}
            \includegraphics[width=\textwidth]{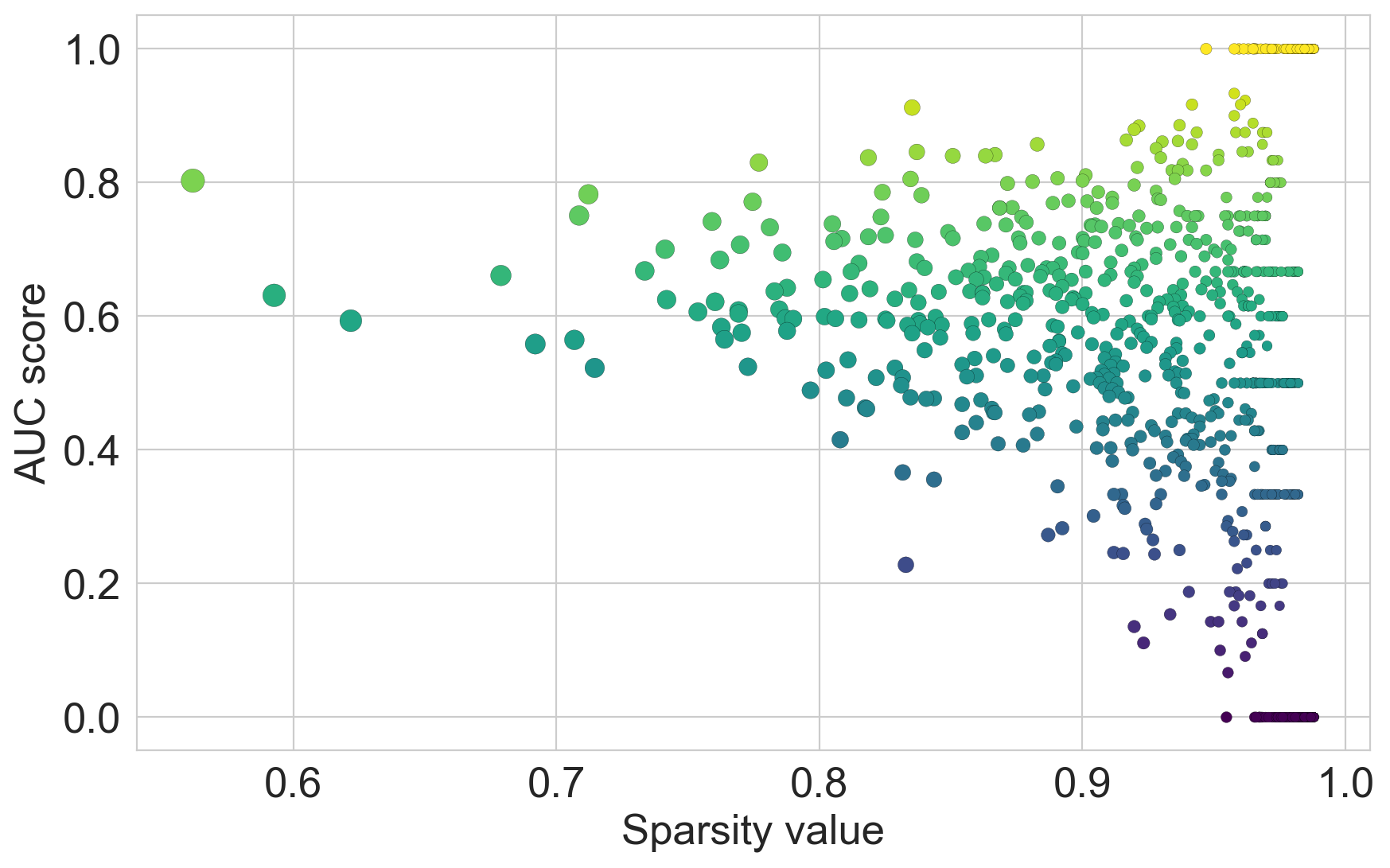}
            \caption{BPR (ML100k)}
            \label{ssfig:figure3}
        \end{subfigure}
    \end{minipage}
    \begin{minipage}[t]{1\textwidth}
        \centering
        \begin{subfigure}[t]{0.24\textwidth}
            \includegraphics[width=\textwidth]{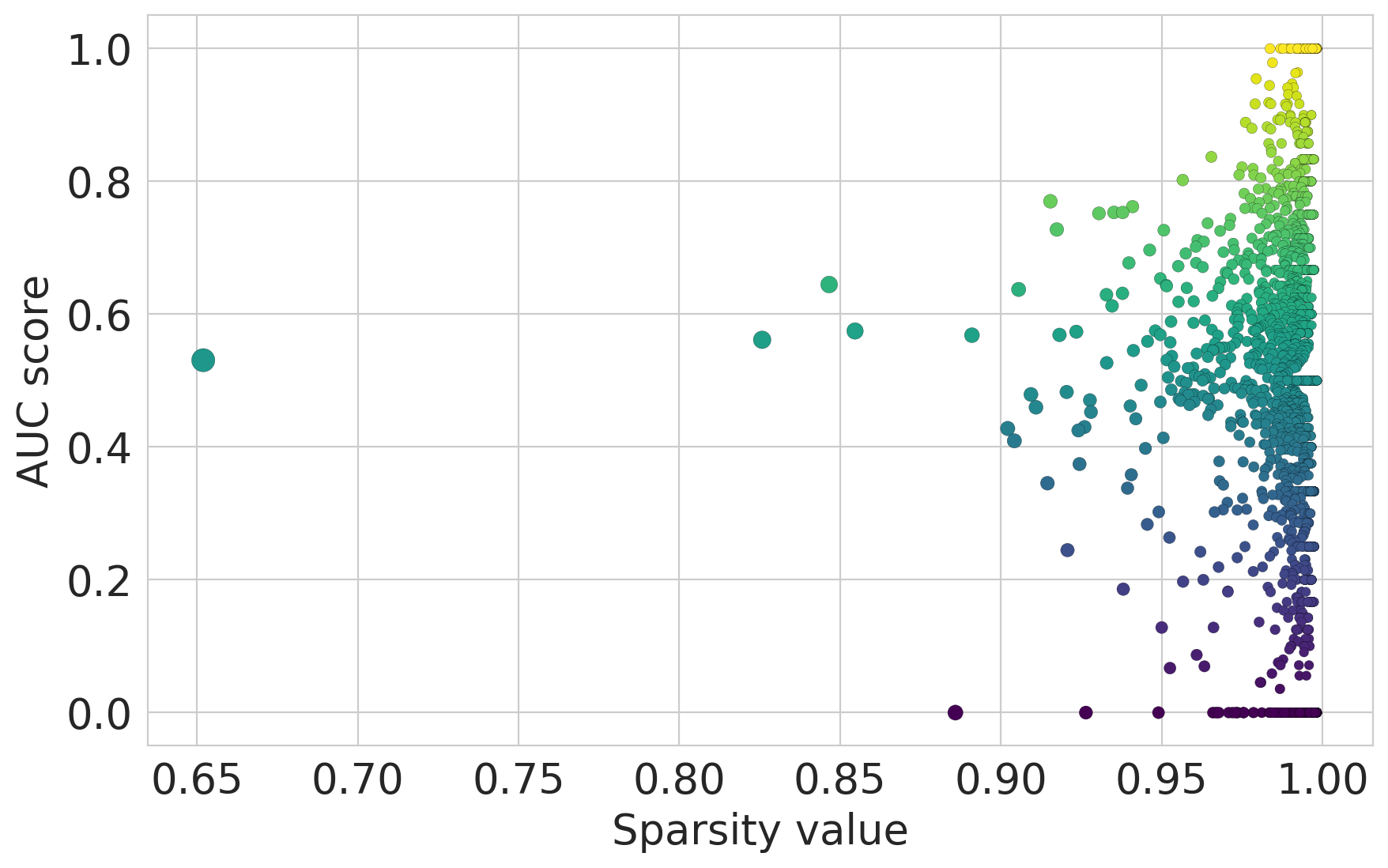}
            \caption{ItemKNN (B-C)}
            \label{sssfig:figure4}
        \end{subfigure}
        \begin{subfigure}[t]{0.24\textwidth}
            \includegraphics[width=\textwidth]{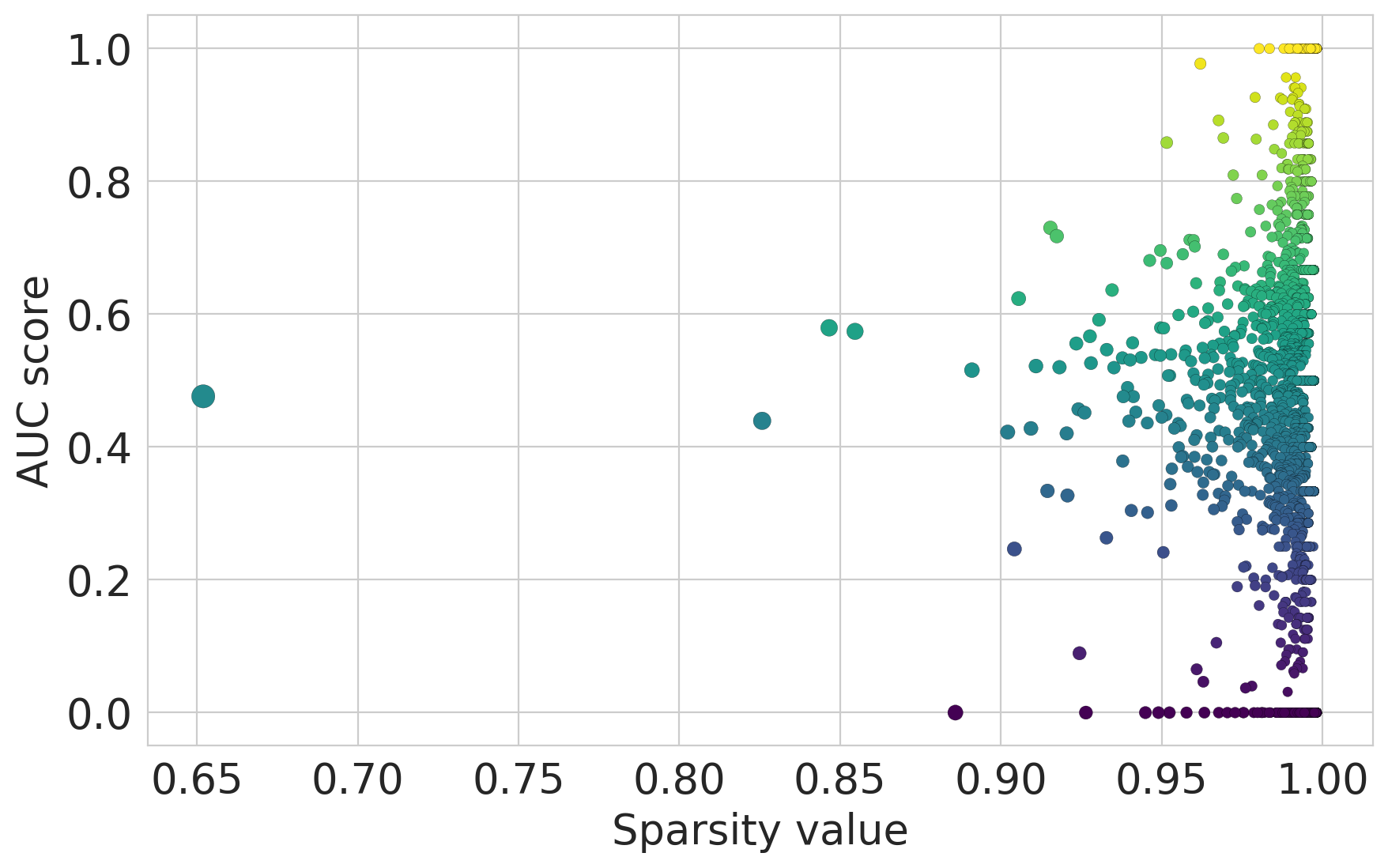}
            \caption{NCF (B-C)}
            \label{ssfig:figure4}
        \end{subfigure}
        \begin{subfigure}[t]{0.24\textwidth}
            \includegraphics[width=\textwidth]{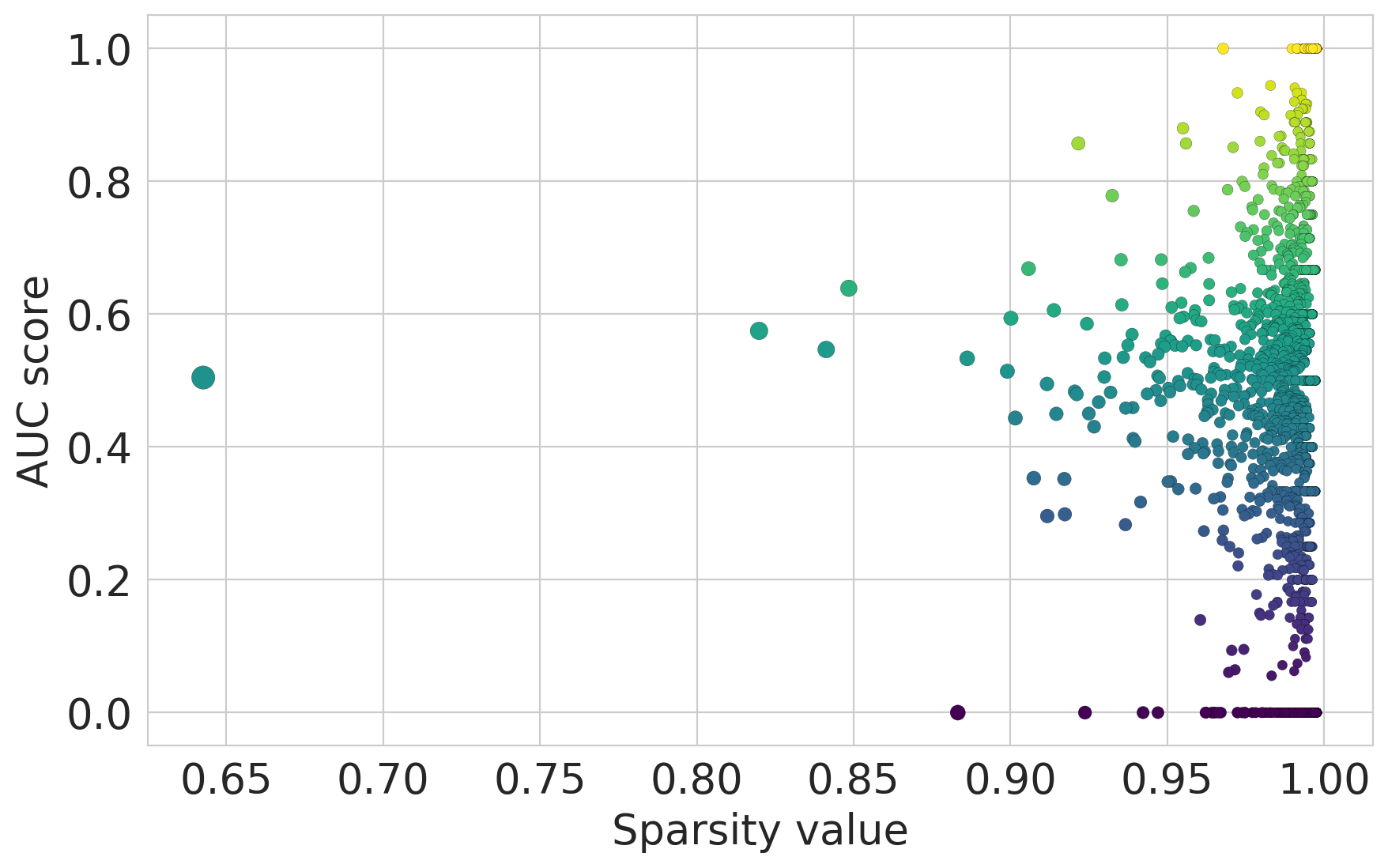}
            \caption{BPR (B-C)}
            \label{sfig:figure4}
        \end{subfigure}
    \end{minipage}
    \caption{AUC vs Sparsity scatter plots for illustrating the performance (measured using AUC- x-axis) for all users in ML1M, ML100k and Book-Crossing (B-C) dataset on three different algorithms.}
    \Description[A scatter plot]{AUC vs Sparsity scatter plots for illustrating the performance (measured using AUC- x-axis) for all users in the ML1M dataset.}
    \label{fig:scatterplot}
\end{figure*}

\subsubsection{Baselines and Models.} Our hybrid framework uses both traditional recommendation systems and LLMs. Thus, we include two different types of recommendation models: (i) Collaborative-filtering based: Neural Collaborative Filtering (NCF)~\cite{he2017neural} as well as       
 ItemKNN~\cite{sarwar2001item}; and  (ii) Learning-to-rank model- Bayesian Personalized Ranking (BPR)~\cite{rendle2012bpr}. 
 While these models identify weak users and generate candidate items, LLMs are further deployed to improve the performance of such users. We use both open (Mixtral-8x-7b-instruct) and closed-sourced (GPT-3.5-turbo)to test the capability of the proposed framework. It is important to note that the Collaborative Filtering models are mostly used to capture the long-term preferences of users. We acknowledge that existing works (refer to Section~\ref{sec:related_work}) have used sequential recommendation models for comparing the performance of LLMs. These works also use recommendation models as candidate retrieval models and then use LLMs to rerank the candidate items. However, sequential models are used to predict the next item according to the recently bought items. We test our framework mainly on long-term user preferences and use collaborative filtering models not only for candidate item retrieval but also for recommending top-k items to strong users. However, we believe that any existing model adopting this retrieval-reranker strategy can adopt our framework. For space constraints, we present an evaluation of our framework only on NCF, ItemKNN and BPR. In line with existing literature~\cite{hou2024large,xu2024prompting}, we design instructions by randomly sampling $20$ rated items to demonstrate the user's preferences to the LLM. Furthermore, existing works do not discuss the responsible adaptation of LLMs, and the underlying task of retrieval-reranker of such models remains consistent and can thus use our framework.

\subsubsection{Implementation details.} For ease of reproducibility, we use the open-source recommendation library RECBOLE~\cite{recbole[2.0]} for implementing all recommendation models and API calls for access to LLMs~\footnote{https://platform.openai.com/docs/api-reference}\footnote{https://www.llama-api.com/}. Each dataset is split into train $(80\%)$, test $(10\%)$ and validation set $(10\%)$. We carefully use the validation set to tune all recommendation models' hyperparameters. For BPR, we search for optimal learning rate in $[5e-5,1e-4,5e-4,7e-4,1e-3,5e-3,7e-3]$ and in $[5e-7,1e-6,5e-6,1e-5,1e-4,1e-3]$ for NCF. Additionally, we use $[64,32,16]$ as  MLP hidden size for all layers and search optimal dropout probability within $[0.0,0.1,0.3]$ for NCF. Two hyperparameter for ItemKNN involve $k$ (neighborhood size) in $[10,50,100,200,250,300,400]$ and $shrink$ (normalization parameter to calculate cosine distance) in $[0.0,0.1,0.5,1,2]$. 
We adopt the protocol presented by a recently released toolkit RGRecSys~\cite{10.1145/3488560.3502192} for evaluating robustness to sub-population using NDCG and AUC. We emphasize that the use of AUC to measure the hardness associated with each user for a given recommendation model is because of the consistency property of AUC. We use the popular CatBoost\footnote{https://github.com/catboost/} library that offers AUC implementation for ranking and also report final NDCG@10 scores. Furthermore, we set the temperature to $0$ in GPT-3.5-turbo to minimize the generation of out-of-list items and hallucinations. However, as per our observations, setting the temperature to $0$ in Mixtral-8x7b-instruct, the model outputs the list in the same order in which it was given input to it. Hence, we set the temperature to $1$ and removed the items which were not originally present in the candidate list. We now discuss our empirical inferences as we conduct experiments following these details. 

\subsection{Empirical Evaluation}
\begin{table*}[t!]
\centering
\caption{Tabular illustration of the overall comparison of results in terms of ranking quality measured using AUC and NDCG@10 for two collaborative filtering based (Neural Collaborative Filtering, ItemKNN) and one learning-to-rank models in comparison to their usage within our framework along with one open-sourced LLM (GPT-3.5-turbo) and one close-sourced LLM (Mixtral-8x7b-instruct).}
\label{tab:Rs+LLMS}
\resizebox{\textwidth}{!}{%
\begin{tabular}{c|cccc|cccc|cccc}
\hline
                                         & \multicolumn{4}{c|}{\textbf{ML1M}}                                                                                                                                                                                                                   & \multicolumn{4}{c|}{\textbf{ML100K}}                                                                                                                                                                                                                 & \multicolumn{4}{c}{\textbf{Book-Crossing}}                                                                                                                                                                                                           \\ \hline
                                         & \multicolumn{1}{c|}{\textbf{AUC}}     & \multicolumn{1}{c|}{\textbf{\begin{tabular}[c]{@{}c@{}}AUC\\  (Weak Users)\end{tabular}}} & \multicolumn{1}{c|}{\textbf{NDCG@10}} & \textbf{\begin{tabular}[c]{@{}c@{}}NDCG@10 \\ (Weak Users)\end{tabular}} & \multicolumn{1}{c|}{\textbf{AUC}}     & \multicolumn{1}{c|}{\textbf{\begin{tabular}[c]{@{}c@{}}AUC\\  (Weak Users)\end{tabular}}} & \multicolumn{1}{c|}{\textbf{NDCG@10}} & \textbf{\begin{tabular}[c]{@{}c@{}}NDCG@10 \\ (Weak Users)\end{tabular}} & \multicolumn{1}{c|}{\textbf{AUC}}     & \multicolumn{1}{c|}{\textbf{\begin{tabular}[c]{@{}c@{}}AUC\\  (Weak Users)\end{tabular}}} & \multicolumn{1}{c|}{\textbf{NDCG@10}} & \textbf{\begin{tabular}[c]{@{}c@{}}NDCG@10 \\ (Weak Users)\end{tabular}} \\ \hline
\textbf{ItemKNN}                         & \multicolumn{1}{c|}{0.47032}          & \multicolumn{1}{c|}{0.23776}                                                              & \multicolumn{1}{c|}{0.66792}          & 0.58226                                                                  & \multicolumn{1}{c|}{0.45616}          & \multicolumn{1}{c|}{0.24778}                                                              & \multicolumn{1}{c|}{0.66792}          & 0.58226                                                                  & \multicolumn{1}{c|}{0.43309}          & \multicolumn{1}{c|}{0.25909}                                                              & \multicolumn{1}{c|}{0.75197}          & 0.65098                                                                  \\ \hline
\textbf{ItemKNN + GPT-3.5-turbo}         & \multicolumn{1}{c|}{\textbf{0.58142}} & \multicolumn{1}{c|}{\textbf{0.51776}}                                                     & \multicolumn{1}{c|}{\textbf{0.82643}} & \textbf{0.70352}                                                         & \multicolumn{1}{c|}{0.59781}          & \multicolumn{1}{c|}{0.51953}                                                              & \multicolumn{1}{c|}{\textbf{0.82643}} & \textbf{0.82598}                                                         & \multicolumn{1}{c|}{\textbf{0.61629}} & \multicolumn{1}{c|}{\textbf{0.49713}}                                                     & \multicolumn{1}{c|}{\textbf{0.86212}} & \textbf{0.77101}                                                         \\ \hline
\textbf{ItemKNN + Mixtral-8x7b-instruct} & \multicolumn{1}{c|}{0.56035}          & \multicolumn{1}{c|}{0.51708}                                                              & \multicolumn{1}{c|}{0.70147}          & 0.69276                                                                  & \multicolumn{1}{c|}{\textbf{0.59972}} & \multicolumn{1}{c|}{\textbf{0.52327}}                                                     & \multicolumn{1}{c|}{0.70147}          & 0.82438                                                                  & \multicolumn{1}{c|}{0.55203}          & \multicolumn{1}{c|}{0.47215}                                                              & \multicolumn{1}{c|}{0.85904}          & 0.76183                                                                  \\ \hline
\textbf{NCF}                             & \multicolumn{1}{c|}{0.47805}          & \multicolumn{1}{c|}{0.22945}                                                              & \multicolumn{1}{c|}{0.78795}          & 0.59801                                                                  & \multicolumn{1}{c|}{0.48311}          & \multicolumn{1}{c|}{0.25182}                                                              & \multicolumn{1}{c|}{0.78795}          & 0.67734                                                                  & \multicolumn{1}{c|}{0.51852}          & \multicolumn{1}{c|}{0.29004}                                                              & \multicolumn{1}{c|}{0.78370}          & 0.66148                                                                  \\ \hline
\textbf{NCF + GPT-3.5-turbo}             & \multicolumn{1}{c|}{\textbf{0.58935}} & \multicolumn{1}{c|}{\textbf{0.52122}}                                                     & \multicolumn{1}{c|}{\textbf{0.80317}} & \textbf{0.71723}                                                         & \multicolumn{1}{c|}{0.60831}          & \multicolumn{1}{c|}{0.50814}                                                              & \multicolumn{1}{c|}{\textbf{0.80317}} & \textbf{0.82603}                                                         & \multicolumn{1}{c|}{\textbf{0.61946}} & \multicolumn{1}{c|}{\textbf{0.50513}}                                                     & \multicolumn{1}{c|}{\textbf{0.88219}} & \textbf{0.78133}                                                         \\ \hline
\textbf{NCF + Mixtral-8x7b-instruct}     & \multicolumn{1}{c|}{0.57211}          & \multicolumn{1}{c|}{0.52100}                                                              & \multicolumn{1}{c|}{0.79178}          & 0.70741                                                                  & \multicolumn{1}{c|}{\textbf{0.61174}} & \multicolumn{1}{c|}{\textbf{0.50903}}                                                     & \multicolumn{1}{c|}{0.79178}          & 0.82306                                                                  & \multicolumn{1}{c|}{0.59901}          & \multicolumn{1}{c|}{0.49897}                                                              & \multicolumn{1}{c|}{0.86254}          & 0.78001                                                                  \\ \hline
\textbf{BPR}                             & \multicolumn{1}{c|}{0.57957}          & \multicolumn{1}{c|}{0.37824}                                                              & \multicolumn{1}{c|}{0.88833}          & 0.73998                                                                  & \multicolumn{1}{c|}{0.51629}          & \multicolumn{1}{c|}{0.17020}                                                              & \multicolumn{1}{c|}{0.88833}          & 0.71944                                                                  & \multicolumn{1}{c|}{0.53310}          & \multicolumn{1}{c|}{0.24426}                                                              & \multicolumn{1}{c|}{0.80405}          & 0.70289                                                                  \\ \hline
\textbf{BPR + GPT-3.5-turbo}             & \multicolumn{1}{c|}{\textbf{0.65397}} & \multicolumn{1}{c|}{\textbf{0.51997}}                                                     & \multicolumn{1}{c|}{\textbf{0.90098}} & \textbf{0.82117}                                                         & \multicolumn{1}{c|}{0.6387}           & \multicolumn{1}{c|}{0.51435}                                                              & \multicolumn{1}{c|}{\textbf{0.90098}} & \textbf{0.82452}                                                         & \multicolumn{1}{c|}{\textbf{0.62145}} & \multicolumn{1}{c|}{\textbf{0.50998}}                                                     & \multicolumn{1}{c|}{\textbf{0.88173}} & \textbf{0.81933}                                                         \\ \hline
\textbf{BPR + Mixtral-8x7b-instruct}     & \multicolumn{1}{c|}{0.64174}          & \multicolumn{1}{c|}{0.51972}                                                              & \multicolumn{1}{c|}{0.89742}          & 0.82998                                                                  & \multicolumn{1}{c|}{\textbf{0.64910}} & \multicolumn{1}{c|}{\textbf{0.52135}}                                                     & \multicolumn{1}{c|}{0.89742}          & 0.81761                                                                  & \multicolumn{1}{c|}{0.61625}          & \multicolumn{1}{c|}{0.50081}                                                              & \multicolumn{1}{c|}{0.87798}          & 0.80284                                                                  \\ \hline
\end{tabular}%
}
\end{table*}

\subsubsection{Comparative analysis.} The \emph{first phase} begins by identifying the inactive users. For this, we calculate the average sparsity of all users in the dataset and identify users above this threshold as inactive users. However, one can use different values for this threshold like~\cite{li2021user} used only top $20\%$ of the sparse users for annotating the inactive users. We then evaluate the AUC score using equation~\ref{eq:p(u)} to measure the performance of RS on all these users. In line with the findings of~\citet{li2021user}, RS performs significantly well on active users as compared to inactive users. For instance-by-instance analysis of every user, we then plot AUC scores against the sparsity index as shown in Fig.~\ref{fig:scatterplot} for all three datasets using three different recommendation algorithms: ItemKNN, NCF, and BPR. While ItemKNN and NCF are collaborative filtering algorithms, the overall scatterplot for BPR shows better AUC scores as compared to the other two algorithms. This is because of the inherent nature of learning-to-rank models like BPR, which rank user preferences better. This figure shows that though RS performs poorly on inactive users on average, not all inactive users receive poor-quality recommendations. We thus use our definition~\ref{def1} to identify such users and mark them as weak. It might be interesting to explore why not all inactive users receive poor performance. We leave this exploratory study for future work. 

For the \emph{second phase}, we design instructions for these weak users using the approach discussed inSection~\ref{sec:methodology}. Our results in Table~\ref{tab:Rs+LLMS} show that LLMs perform significantly well on these users. Using LLM and base RS models yields the best results for all three datasets. Our results show improvement in both AUC and NDCG@10 for weak users, thus demonstrating improved robustness to the sub-population of weak users. This further leads to an overall improved ranking quality. We also highlight that previous works like~\cite{yue2023llamarec,hou2024large} show that close-source models perform much better than open-source. However, understanding the properties of users for which LLMs inherently perform well (like we provide a mechanism of finding weak users) and responsibly assigning tasks to large models improve performance using open-source as well as closed-source models. Our results show that Mixtral-8x-7b-instruct can perform almost equally well on weak users in all the datasets and base models. Furthermore, as observed for the ML100k dataset, this open-source model can outperform GPT-3.5-turbo when evaluated on AUC. It should be noted that AUC mainly evaluates the discriminatory ability of model to rank positive items over negative and NDCG focuses on the user's satisfaction with the ranked list, considering both relevance and position. The reason for this can be associated with dataset sparsity. As shown in Table~\ref{tab:my-table}, the sparsity of the ML100k dataset is lesser $(\approx 93\%)$  compared to the other two datasets. While Mixtral is a good choice when datasets are small and more dense, GPT-3.5-turbo performs well for extremely sparse datasets. Yet, it is important to note that in either case, the margin of the performance of both these LLMs is not significant, and thus, even open-source models can give a comparable performance. Nevertheless, usage of GPT model yields best NDCG@10 scores for all datasets.
\begin{figure*}
    \centering
    \begin{subfigure}[b]{0.3\textwidth}
        \centering
        \includegraphics[width=\textwidth]{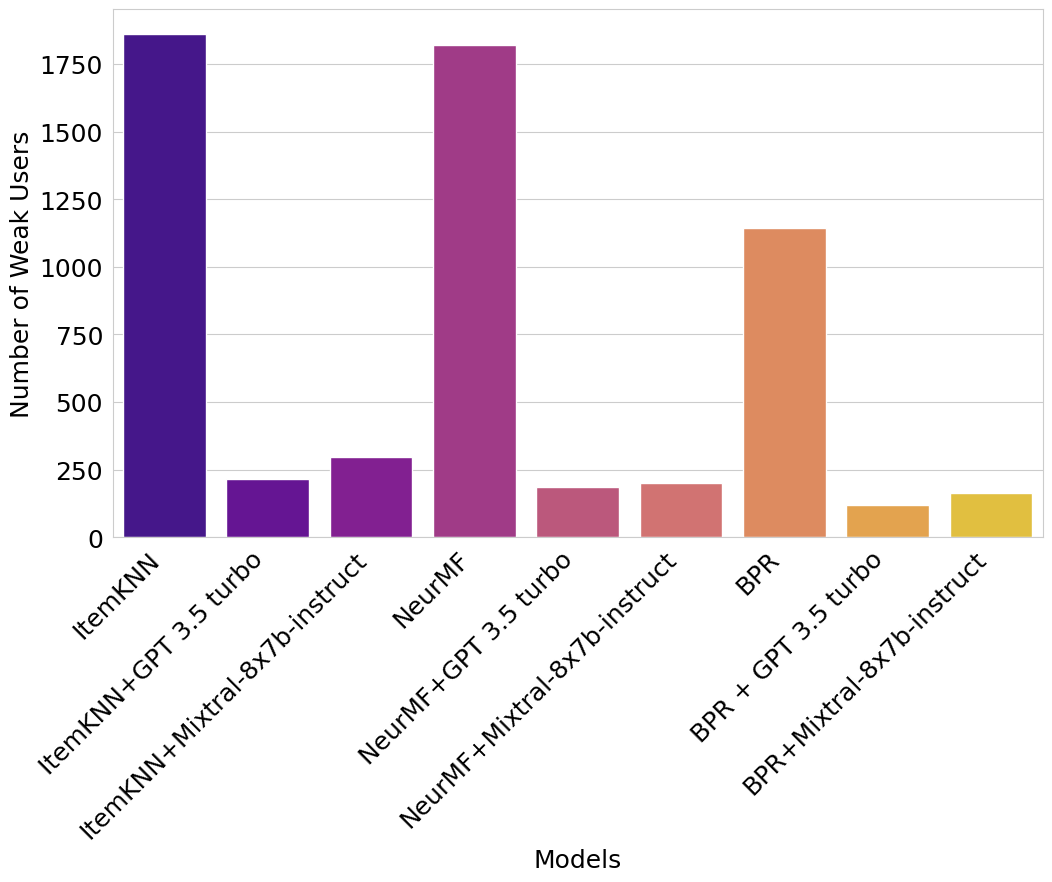}
        \caption{ML1M}
        \label{fig:figure1}
    \end{subfigure}
    \hfill
    \begin{subfigure}[b]{0.3\textwidth}
        \centering
        \includegraphics[width=\textwidth]{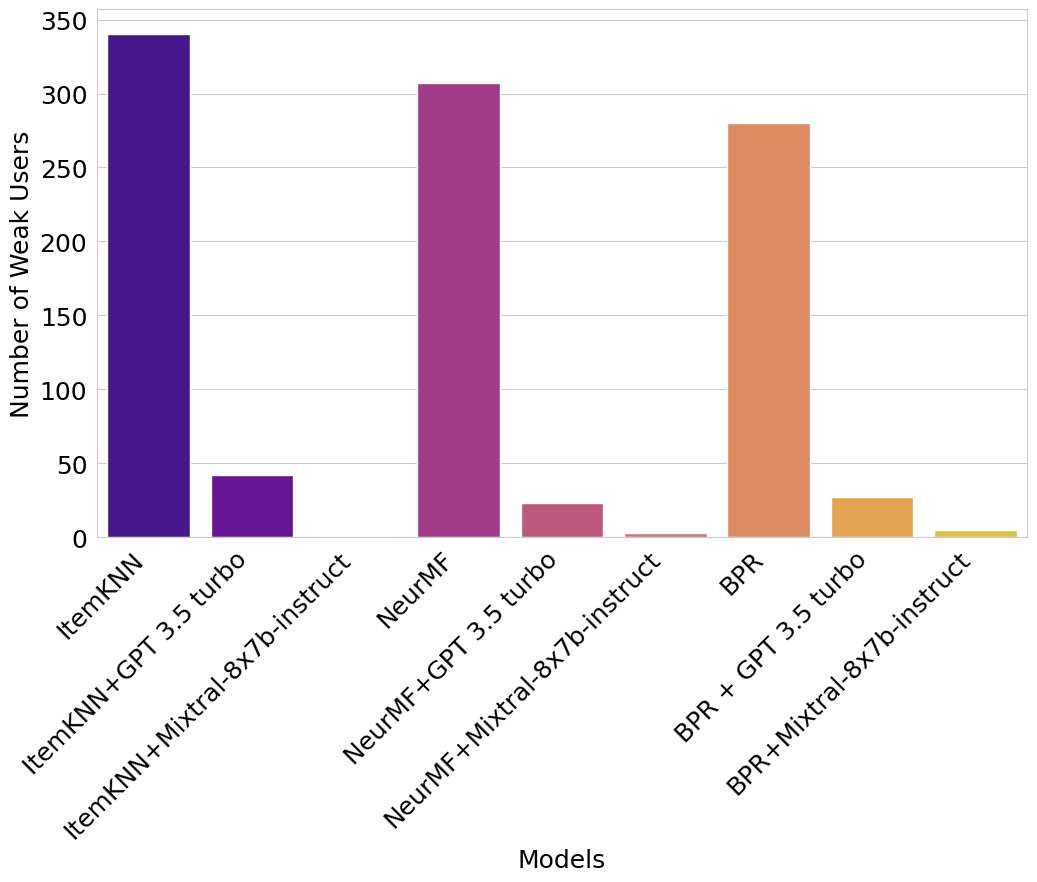
        }
        \caption{ML100k}
        \label{ffig:figure2}
    \end{subfigure}
    \hfill
    \begin{subfigure}[b]{0.3\textwidth}
        \centering
        \includegraphics[width=\textwidth]{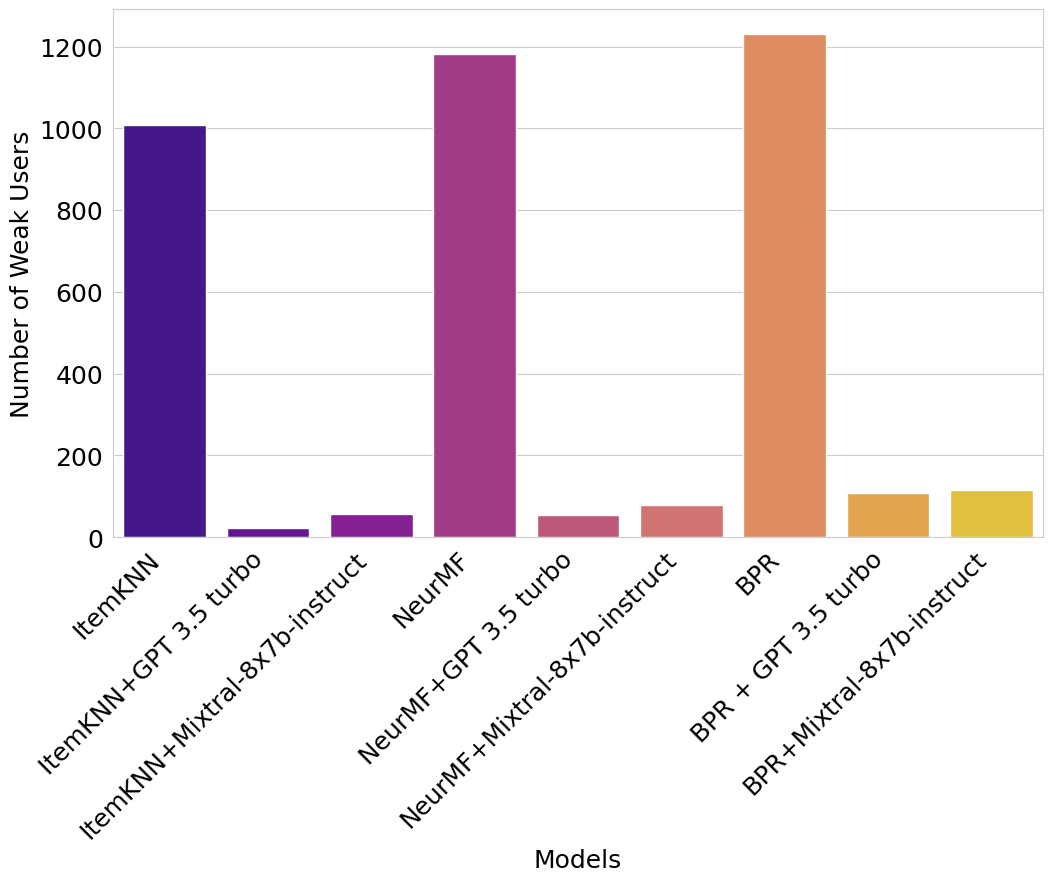}
        \caption{Book-Crossing}
        \label{fig:figure2}
    \end{subfigure}
    \caption{Comparative analysis of reduction in the count of weak users }
    \Description[Reduction in weak user count]{Comparative analysis of reduction in the count of weak users }
    \label{fig:barplot}
\end{figure*}

\subsubsection{Reduction in weak user count.} For analyzing the variations in the count of weak users, we counted a number of weak users identified in the first phase whose interactions were contextualized and given to LLMs as the base for comparison with LLMs using the same threshold $t_p$. We evaluated AUC on the rankings obtained by LLM for these users. It was noted that a few users continued being hard even for LLM if the AUC lied below $t_p$. Fig.~\ref{fig:barplot} shows that when used with large models, the count of weak users in recommendation systems drops significantly. In highly sparse datasets like that of Book-Crossing and ML1M, GPT-3.5-turbo reduced the number of weak users by $\approx87\%$ and Mixtral-8x7b-instruct by $\approx85\%$. On the contrary, when the dataset is dense like ML100k, Mixtral-8x7b-instruct can reduce the count by $\approx99\%$ and the closed-source model by $\approx88\%$. While the reduction ability of GPT-3.5-turbo remains consistent over all datasets, the open-source models yield better performance for less sparse datasets yet improve the robustness of RS to sub-populations. 

In addition, we noted that a single query takes $\approx8$ seconds in GPT-3.5-turbo and $\approx11$ seconds in Mixtral-8x7b-instruct. This shows each user query's high processing times and inference latency. Thus, it is crucial to use these models responsibly by identifying what they are good at. Considering the example of the smallest dataset of ML100k, which consists of $943$ users. Our strategy for identifying weak users results in only $330$ weak users (worst case by ItemKNN), which leads to an overhead of $2,640$ seconds in using GPT-3.5-turbo in addition to the training time of base RS models, which is significantly less than the $7,544$ seconds if used for all the users. 

\section{Discussion}
In this work, we implemented a novel approach for responsible adaptation of LLMs for ranking tasks.
As suggested by~\citet{doi:10.1126/science.adf6369}, the involvement of AI in high-stake decision-based applications (like ranking models for job recommendations) requires instance-by-instance evaluation instead of aggregated metrics for designing responsible AI models. Our results in Fig.~\ref{fig:scatterplot} show that many inactive users receive recommendations of poor quality by traditional RS. Some inactive users still receive recommendations comparable to the active users, but this might be due to high similarity scores with active users that existing models can still capture their preferences effectively. We leave this as an exploratory study for future work. However, the overall performance scores on active users remain better than those of inactive users. Building upon these weak instances, our framework emphasizes on instance-by-instance evaluation of users. While we group users based on activity and then evaluate the performance of inactive users, our approach pinpoints to weak users whose preferences remain hard for traditional RS to capture effectively. We believe that our framework inherently addresses the issue of group fairness. Though we do not group users based on demographics, our framework can also be extended to these scenarios where instead of activity, user demographics can be used to group users and then within the marginalized groups, the interaction histories of users who receive poor performance can be contextualized and given to LLM. However, irrespective of the demographics, our framework mainly addressed the robustness to data sparsity and sub-population of weak users, which inherently tackles the fairness issue. 

This framework further helped reduce the number of queries which needed to be given to the LLM. Since most existing works (refer to Section~\ref{sec:related_work}) randomly select a few users out of all the users present in the dataset to evaluate the performance of LLMs, our framework provides a systematic way of selecting users for which LLMs can be used. Leveraging the capabilities of LLMs for weak users, our work emphasizes the importance of low-cost traditional RSs as well. We also observed that in some cases (Fig.~\ref{fig:barplot}), LLM might not be able to perform well on every weak user. This opens up new research opportunities for understanding similarities and differences within the identified weak users on which LLM does and does not perform well. Further, one can think of various prompting strategies to prompt the model to capture the preferences of extremely weak users effectively. Past works have developed various prompting strategies, which can all be tested to observe which strategies remain effective for which types of users. Nevertheless, the main goal of this paper remains to emphasise the importance of responsible adaptation of LLMs by strategically selecting tasks for which these models inherently perform well. It is also important to note that for weak users, we still obtain candidate items from traditional RS, as has been done in most past works (refer to Section \ref{sec:related_work}). This helped in reducing the candidate set from thousands of unrated items to a few, which were given to the LLM to then rank. While this approach ensures that the results obtained by RS for weak users are utilized for generating candidate items instead of discarding such results directly, thus maximizing the usage of RSs even for weak users, it has a limitation. Traditional RSs perform worse on these users, and the candidate items might not capture the true preference of weak users. When we give these candidate items to LLM for ranking, the results might deviate further from true preferences. This issue can be the one reason that LLMs might not perform well on all weak users. One can, thus, further investigate the relation of candidate items to the performance of LLMs on certain users. If the candidate items are already non-preferred items by users, LLM might inherently find it difficult to perform well. 

Our work, thus, represents a foundational step towards responsibly adapting LLMs while emphasizing the importance of traditional models, particularly focusing on addressing the challenges posed by sub-populations with sparse interaction histories. Our instance-by-instance evaluation approach, inspired by the imperative highlighted by recent studies in high-stakes decision-based AI applications, underscores the necessity of a nuanced understanding of individual user needs and preferences. While our framework emphasizes the importance of leveraging traditional recommendation systems alongside LLMs, we acknowledge the need to further explore the performance variations among weak users and the impact of candidate item selection on LLM effectiveness. Moving forward, our work lays the groundwork for continued research into refining the adaptation of LLMs, ensuring their responsible deployment across diverse user populations and application scenarios.

\section{Conclusion}
In this paper, we presented a hybrid framework that aims to improve the robustness of RS to sub-populations by leveraging LLMs. Our approach first utilized user activity to identify inactive users and then measured the performance of various RSs on these users to pinpoint weak users on which RSs find it hard to perform well. In doing so, this paper represents a novel stride towards improving robustness to sub-populations (irrespective of sensitive attributes) in RSs through efficient and responsible adaptation of LLMs without requiring any fine-tuning. While we evaluated our framework for the long-term preference of users using various collaborative filtering and learning-to-rank models, our model can be extended to various recommendation models. Our work particularly examines the importance of evaluating various user properties in the responsible adaptation of generative models within the recommendation domains. This paper opens numerous research directions for further exploration including developing prompting strategies for extremely weak users on which LLMs can not perform well. We further aim to explore other factors that can aid in the responsible adaptation of large models and improve the robustness and fairness of the challenges within recommendation systems.
\bibliography{sample-base}
\end{document}